\documentclass{osa-article}

\journal{osac}
\usepackage{algorithm,algorithmic}
\usepackage{subcaption}
\usepackage{rotating}
\graphicspath{{images/}}
\usepackage{tikz}
\usepackage{comment}

\articletype{Research Article}

\begin{document}

\title{A semi-implicit relaxed Douglas-Rachford algorithm (sir-DR) for Ptychograhpy}

\author{Minh Pham,\authormark{1,*} Arjun Rana,\authormark{2} Jianwei Miao \authormark{2}, and Stanley Osher\authormark{1}}

\address{\authormark{1}Department of Mathematics, University of California, Los Angeles, CA 90095, USA\\
\authormark{2}Department of Physics \& Astronomy and California NanoSystems Institute, University of California, Los Angeles, CA 90095, USA

\authormark{*}minhrose@math.ucla.edu
}



\begin{abstract}
Alternating projection based methods, such as ePIE and rPIE, have been used widely in ptychography. However, they only work well if there are adequate measurements (diffraction patterns); in the case of sparse data (i.e. fewer measurements) alternating projection underperforms and might not even converge. In this paper, we propose semi-implicit relaxed Douglas Rachford (sir-DR), an accelerated iterative method, to solve the classical ptychography problem. Using both simulated and experimental data, we show that sir-DR improves the convergence speed and the reconstruction quality relative to ePIE and rPIE. Furthermore, in certain cases when sparsity is high, sir-DR converges while ePIE and rPIE fail. To facilitate others to use the algorithm, we post the Matlab source code of sir-DR on a public website (www.physics.ucla.edu/research/imaging/sir-DR). We anticipate that this algorithm can be generally applied to the ptychographic reconstruction of a wide range of samples in the physical and biological sciences. 
\end{abstract}

\section{Introduction}
Since the first experimental demonstration in 1999 \cite{Miao1999}, coherent diffraction imaging (CDI) through directly inverting far-field diffraction patterns to high-resolution images has been a rapidly growing field due to its broad potential applications in the physical and biological sciences \cite{Miao2015, Gaffney2007, Robinson2009, Pfeiffer2018}. A fundamental problem of CDI is the phase problem, that is, the diffraction pattern measured only contains the magnitude, but the phase information is lost. In the original demonstration of CDI, phase retrieval was performed by measuring the diffraction pattern from a finite object. If the diffraction intensity is sufficiently oversampled \cite{Miao1998}, the phase information can be directly retrieved by using iterative algorithms \cite{Fienup1982, Luke, Rodriguez2013, Marchesini2007, Shechtman2015}. Ptychography, a powerful scanning CDI method, relieves the finite object requirement by performing 2D scanning of an extended relative to an illumination probe and measuring multiple diffraction patterns with each illumination probe overlapped with its neighboring ones \cite{Rodenburg2007,Thibault08}. The overlap of the illumination probes not only reduces the oversampling requirement, but also improves the convergence speed of the iterative process. By taking advantage of ever-improving computational power and advanced detectors, ptychography has been applied to study a wide range of samples using both coherent x-rays and electrons \cite{Miao2015, Pfeiffer2018, Dierolf, Holler2017, Gallagher2017, Maiden15, Shapiro2014, Gardner2017, Marrison2013, Deng2018, Gao2017, Jiang2018}. More recently, a time-domain ptychography method was developed by introducing a time-invariant overlapping region as a constraint, allowing the reconstruction of a time series of complex exit wave of dynamic processes with robust and fast convergence \cite{Lo2018}.

Algorithms for ptychography have been studied exhaustively in theory and practice. The majority following non-convex optimization approaches \cite{Hesse,Alfonso,Bian:15}, while a few follow convex relaxation \cite{Horstmeyer}. In recent years, powerful ptychographic algorithms have been developed to handle partial coherence \cite{Thibault13}, solve for the probe \cite{Fienup08,Thibault08,Maiden09}, correct positioning errors \cite{Maiden12,Zhang13,Tripathi14}, reduce noise \cite{Thibault12,Godard12}, and deal with
multiple scattering \cite{Maiden12b,Tsai16}. 

These iterative algorithm can be generally divided into three classes: i) the conjugate gradient (CG) \cite{Fienup08,Tripathi14}, ii) extended ptychography iterative engine (ePIE) \cite{Maiden09}, and iii) difference map (DM)  \cite{Thibault08}, whereas the last two have a close relationship. ePIE is an alternating projection algorithm, while DM is built on both projection and reflection which is believed to provide a momentum to speed up the convergence. The relaxed average alternating reflection (RAAR) \cite{Luke} is a relaxation of DM and has been shown to be effective in phase retrieval \cite{Marchesini16}. All algorithms except ePIE take a global approach, i.e. using the entire collection of diffraction patterns to perform an update of the probe and object in each iteration. In contrast, ePIE goes through the measured data sequentially to refine the probe and object. However, ePIE has a slow convergence due to the step size restriction which may cause divergence if violated. To fix this issue, rPIE was proposed, in which regularization is used for stability \cite{Maiden:17}. The significant results also show that rPIE obtains a larger field of view (FOV) than ePIE. 

In this paper, we show that DM and RAAR can also be implemented locally, similarly to ePIE. We then apply non-convex optimization tools to improve the robustness, convergence and FOV of the ptychography reconstruction. The proposed algorithm incorporates two techniques. The first modifies the update of the probe and object as the algorithm iterates via semi-implicit method or Proximal Mapping. The second technique is the implementation of relaxed Douglas Rachford, a generalized version of DM and RAAR, on the local scale.

\section{The proposed algorithm}
Given N measured diffraction patterns at N positions, the ptychographic algorithm aims to find a 2D object O and probe P that satisfy the overlap constraint and the Fourier magnitude constraint
\begin{align}
	|\mathcal{F} (P O_n) | = \sqrt{I_n} \quad \mbox{for} \quad n=1,..,N. 
\end{align}
Where $O_n$ is the object at the $n^{th}$ scan position. Here, we omit the spatial variables for a simple notation and use the notations $P, \, O_n$ and $I_n$ for both continuous and discrete cases. The absolute value, multiplication, division, conjugate, and square root operators are applied element-wise on $P, \, O_n$ and $I_n$ which represent 2D complex matrices of the same size in the discrete case. We can argue that $O_n$ is the object restricted to a sub-domain $\Omega_n$. The overlap constraint can be mathematically interpreted as
\begin{align}
	O(x+r_n) = O_n(x) \quad \mbox{if} \quad r_n+x \in \Omega_n \quad \mbox{for} \quad n=1,...,N
\end{align}
where $\{r_n\}_{n=1}^N$ are displacement vectors. In short notation, we write $O_n = O|_{\Omega_n}$ to imply the object is restricted to sub domain $\Omega_n$. The equivalent constraint in the discrete case is the agreement between the sub-matrix of $O$ and $O_n$. We find a better representation of the problem by introducing the exit wave variable $\Psi = PO$. By denoting the Fourier measurement constraint set $T$ and the overlap object constraint set $S$, we have
\begin{align*}
	\mathcal{T} &:= \big\{ \Psi = \{\Psi_n\}_{n=1}^N: | \mathcal{F} \Psi_n| = \sqrt{I_n} \mbox{ for } n=1,...,N \big\} \\
  	\mathcal{S} &: = \big\{ \Psi = \{\Psi_n\}_{n=1}^N: \exists P, O \mbox{ s.t. } \Psi_n = PO_n  \mbox{ for } n=1,...,N \}\big\}.
\end{align*}
Then we write the ptychography problem in a minimization fashion
\begin{align}\label{min:ePIE}
	\min_{ \Psi} \quad i_{\mathcal{S}}(\Psi) + i_{\mathcal{T}}(\Psi)
\end{align}
where $i_{\mathcal{S}}(\Psi)$ and $i_{\mathcal{T}}(\Psi)$ are the indicator functions of sets $\mathcal{S}$ and $\mathcal{T}$ respectively, defined as
\begin{align}
    i_{\mathcal{S}} (\Psi) = \begin{cases}
        0 & \Psi \in \mathcal{S} \\
        \infty & \text{otherwise}
    \end{cases}
\end{align}
To solve Eq. (\ref{min:ePIE}), an alternating projection method is proposed. At each iteration, we select a random position $n$ and update $\Psi_n$
\begin{align}
	\Psi_n' &= \Pi_{\mathcal{T}} ( \Psi_n^k ) = \mathcal{F}^{-1} \big( \sqrt{I_n}  \arg (\mathcal{F} \Psi_n^k) \big) \\
    \{ P^{k+1},O_n^{k+1} \} &= \mathop{\mathrm{argmin}}_{P,O_n} \frac{1}{2} \| PO_n - \Psi_n' \|^2 \label{eqn:min_OP1}\\
    \Psi_n^{k+1} &= P^{k+1} O_n^{k+1}
\end{align}
The Frobenius norm is used in this minimization problem and entire paper unless a different norm is specified. The minimization of Eq. ($\ref{eqn:min_OP1}$) is difficult due to instability. One way to solve this non-convex problem is to minimize each variable while fixing the other ones.
\begin{equation} \label{eqn:OP_AD}
\begin{aligned}
	O_n^{k+1} &=  \mathop{\mathrm{argmin}}_{O_n} \frac{1}{2} \| P^k O_n - \Psi_n' \|^2 = \frac{\Psi_n'}{P^k} \\
    P^{k+1} &=  \mathop{\mathrm{argmin}}_{P} \frac{1}{2} \| P O_n^{k+1} - \Psi_n' \|^2 = \frac{\Psi_n'}{O_n^{k+1}}
\end{aligned}
\end{equation}
This approach is unstable because of the division. A cut-off method is used to avoid the divergence and zero-division. A modification is recommended by adding a penalizing least square error term (i.e. regularizer)
\begin{align} \label{eqn:min_OP2}
	\{ P^{k+1},O_n^{k+1} \} &= \mathop{\mathrm{argmin}}_{P,O_n} \frac{1}{2} \| P O_n - \Psi_n' \|^2 + \frac{1}{2 s} \| P-P^k \|^2 +\frac{1}{2t} \|O_n-O_n^k\|^2
\end{align}
The idea of regularization appears throughout the literature such as proximal algorithms \cite{Bertsekas,Parikh}. Eq. (\ref{eqn:min_OP2}) is more reliable to solve than Eq. (\ref{eqn:min_OP1}) but is still very expensive since the variables are coupled. $P^{k+1}$ and $O_n^{k+1}$ can be solved via a Backward-Euler system derived from Eq. (\ref{eqn:min_OP2}).
\begin{equation} \label{eqn:OP_BE}
\begin{aligned}
	O_n^{k+1} &= O_n^k -t \overline{P^{k+1}} \big( P^{k+1} O_n^{k+1} - \Psi_n' \big) \\
    P^{k+1} &= P^k - s \overline{O_n^{k+1}} \big( P^{k+1} O_n^{k+1} - \Psi_n' ) 
\end{aligned}
\end{equation}
ePIE proposes a simple approximation by linearizing the system so that it can be solved sequentially. 
\begin{equation}\label{eqn:ePIE}
\begin{aligned}
	\quad O_n^{k+1} &= O_n^k -t \overline{P^k} \big( P^k O_n^k - \Psi_n' \big) \\
    \quad P^{k+1} &= P^k - s \overline{O_n^{k+1}} \big( P^k O_n^{k+1} - \Psi_n' )
\end{aligned}
\end{equation}
The system is solved by alternating direction methods (ADM)  \cite{Marchesini3}. The remaining part is to choose appropriate step sizes $t$ and $s$ to ensure stability. ePIE suggests $t = \beta_O/ \|P^k\|_{\max}^2$ and $s = \beta_P/ \|O^{k+1}\|_{\max}^2$ where $\beta_O, \; \beta_P \in (0,1]$ are normalized step sizes. The max matrix norm is the element-wise norm, taking the maximum in absolute values of all elements in the matrix. The final version of ePIE is given by
\begin{equation}
\begin{aligned}
	O_n^{k+1} &= O_n^k - \beta_O \frac{\overline{P^k} \big( P^k O_n^k - \Psi_n' \big)}{\|P^k\|_{\max}^2}  \\
    P^{k+1} &= P^k - \beta_P \frac{\overline{O_n^{k+1}} \big( P^k O_n^{k+1} - \Psi_n' )}{\|O_n^{k+1}\|_{\max}^2} 
\end{aligned}
\end{equation}

We will exploit the structure of Eq. (\ref{eqn:OP_BE}) to give a better approximation.

\medskip

\subsection{A semi-implicit algorithm}
We replace the minimization of Eq. (\ref{eqn:min_OP2}) by two steps
\begin{equation}
\begin{aligned}
	\mbox{Step 1:} \quad O_n^{k+1} &= \mathop{\mathrm{argmin}}_{O_n} \frac{1}{2} \| P^k O_n - \Psi_n' \|^2 + \frac{1}{2t} \| O_n - O_n^k \|^2 \\
    \mbox{Step 2:} \quad P^{k+1} &= \mathop{\mathrm{argmin}}_{P} \frac{1}{2} \| P O_n^{k+1} - \Psi_n' \|^2 + \frac{1}{2s} \| P - P^k \|^2 
\end{aligned}
\end{equation}
This results in a  better approximation to the linearized system of Eq. (\ref{eqn:ePIE}) and simpler than the Backward-Euler Eq. (\ref{eqn:OP_BE})
\begin{equation} \label{eqn:OP_semi_BE}
\begin{aligned}
	O_n^{k+1} &= O_n^k -t \overline{P^k} \big( P^k O_n^{k+1} - \Psi_n' \big) \\
    P^{k+1} &= P^k - s \overline{O_n^{k+1}} \big( P^{k+1} O_n^{k+1} - \Psi_n' )
\end{aligned}
\end{equation}
In this uncoupled system, we can derive a closed form solution for each sub-problem.
\begin{equation}
\begin{aligned}
	O_n^{k+1} &= \frac{O_n^k + t \overline{P^k} \Psi_n'}{1 + t |P^k|^2} \\
    P^{k+1} &= \frac{p^k + s \overline{O_n^{k+1}} \Psi_n'}{1 + s|O_n^{k+1}|^2}
\end{aligned}
\end{equation}
By choosing the step sizes $s$ and $t$ as in the ePIE algorithm and normalizing the parameters $\beta_O$ and $\beta_P$, we obtain
\begin{equation} \label{eqn:OP_sBE}
\begin{aligned}
	O_n^{k+1} &= \frac{(1-\beta_O)\|P^k\|_{\max}^2 O_n^k + \beta_O \overline{P^k} \Psi_n' }{(1-\beta_O)\|P^k\|_{\max}^2 + \beta_O |P^k|^2} \\
    P^{k+1} &= \frac{(1-\beta_P)\|O_n^{k+1}\|_{\max}^2 P^k + \beta_P \overline{O_n^{k+1}} \Psi_n' }{(1-\beta_P)\|O_n^{k+1}\|_{\max}^2 + \beta_P |O_n^{k+1}|^2} 
\end{aligned}
\end{equation}
This formula can be explained as a weighted average between the previous update $O_n^k$ and $\displaystyle \frac{\Psi_n^k}{P^k}$. The object update is similar to the rPIE algorithm when rewriting it as
\begin{align}
	O_n^{k+1} = O^k + \beta_O  \frac{ \overline{P^k} \big( \Psi_n' - \Psi_n^k \big) } {(1-\beta_O)\|P^k\|_{\max}^2 + \beta_O |P^k|^2}
\end{align}
The difference is rPIE does not have the parameter $\beta_O$ in front of the fraction. i.e. rPIE has a larger step size than sir-DR. This helps converge faster but might also get trapped in local minima. The regularization (weighted average) in sir-DR is more mathematically correct and enhances the algorithm's stability. 

In the next section, we apply the Douglas-Rachford algorithm to solve for the exit wave $\Psi$.

\subsection{The relaxed Douglas-Rachford algorithm}
The Douglas-Rachford algorithm was originally proposed to solve the heat conduction problem  \cite{Douglas}, which represents a composite minimization problem
\begin{align}
	\min_{\Psi} \quad f(\Psi) + g(\Psi)
\end{align}
The iteration consists of
\begin{align}
	\Psi^{k+1} = \Psi^k + \mathrm{prox}_{tf} \big( 2\; \mathrm{prox}_{tg} (\Psi^k) - \Psi^k \big) - \mathrm{prox}_{tg} (\Psi^k)
\end{align}
Over the past decades, this accelerated convex optimization algorithm has been exhaustively studied in both theory and practice with many applications  \cite{Tseng1991,Lions1979,Eckstein1992,nesterov2013,Tseng,bubeck}.Here we apply the algorithm to the ptychographic phase retrieval. Note that the Douglas-Rachford algorithm reduces to Difference Map (DM) when $f=i_{\mathcal{T}}$ and $g=i_{\mathcal{S}}$ are characteristic functions of constraint sets $\mathcal{T}$ and $\mathcal{S}$, respectively
\begin{align}
	\Psi^{k+1} = \Psi^k + \Pi_{\mathcal{T}} \big( 2\; \Pi_{\mathcal{S}} (\Psi^k) - \Psi^k \big) - \Pi_{\mathcal{S}} (\Psi^k)
\end{align}
We realize that the reflection operator $2\Pi_{\mathcal{S}}-I$ helps to accelerate the convergence in the convex case and escape local minima in the non-convex case. However this momentum, caused by reflection, might be too large and can lead to over-fitting. Therefore, we relax the reflection by introducing the relaxation parameter $\sigma \in [0,1]$
\begin{align}
    \Psi^{k+1} = \mathrm{prox}_{tf} \Big( (1+\sigma)\; \mathrm{prox}_{tg} (\Psi^k) - \sigma\Psi^k \Big) + \sigma \Big(\Psi^k - \mathrm{prox}_{tg} (\Psi^k) \Big)
\end{align}
Since the experimental measurements are contaminated by noise, a direct projection of measurement constraint is not an appropriate approach. We thus relax the Fourier magnitude constraint by a least square penalty
\begin{align} \label{eqn:relax_min}
	\min_{ \Psi_n } \quad \sum_{n=1}^N \| |\mathcal{F} \Psi_n | - \sqrt{I_n} \|^2 +i_{\mathcal{S}}(\Psi_n)
\end{align}
Recall that $\mathrm{prox}_{tf}(\Psi)$ has a closed form solution
\begin{align}
	\mathrm{prox}_{tf}(\Psi^k) &= \mathop{\mathrm{argmin}}_{\Psi} \frac{1}{2} \| | \mathcal{F} \Psi| - \sqrt{I_n} \|^2 + \frac{1}{2t} \| \Psi - \Psi^k \|^2 \nonumber \\
    &= \frac{\Psi^k + t \mathcal{F}^{-1} \Big[\sqrt{I_n}  \arg(\mathcal{F}\Psi) \Big] }{1 + t} \nonumber \\
    &= (1-\tau) \Psi^k + \tau \mathcal{F}^{-1} \Big[\sqrt{I_n}  \arg(\mathcal{F}\Psi) \Big]
\end{align}
where $\tau = t/(1+t) \in (0,1)$ exclusively is the normalized step size. Combining this result with DM, we obtain
\begin{align}
	\Psi^{k+1} &=  (1-\tau) \Big( (1+\sigma)\pi_{\mathcal{S}} (\Psi^k) - \sigma \Psi^k \Big) +\tau \Pi_{\mathcal{T}} \Big( (1+\sigma)\pi_{\mathcal{S}} (\Psi^k) - \sigma\Psi^k \Big) + \sigma \Big( \Psi^k - \Pi_{\mathcal{S}}(\Psi^k) \Big)  \nonumber\\
    &= \tau \Big( \sigma \Psi^k + \Pi_{\mathcal{T}} \big( (1+\sigma)\pi_{\mathcal{S}} (\Psi^k) - \sigma \Psi^k \big) \Big) + \Big(1-\tau(1+\sigma) \Big) \Pi_{\mathcal{S}}(\Psi^k) 
\end{align}
When we let $\beta = 1-\tau$ and $\sigma=1$, the update reduces to RAAR. Therefore, we show that relaxed Douglas-Rachford is a generalized version of RAAR. We now move to our main algorithm.

\subsection{The sir-DR algorithm}
\begin{figure}
    \centering
    \includegraphics[width=13cm]{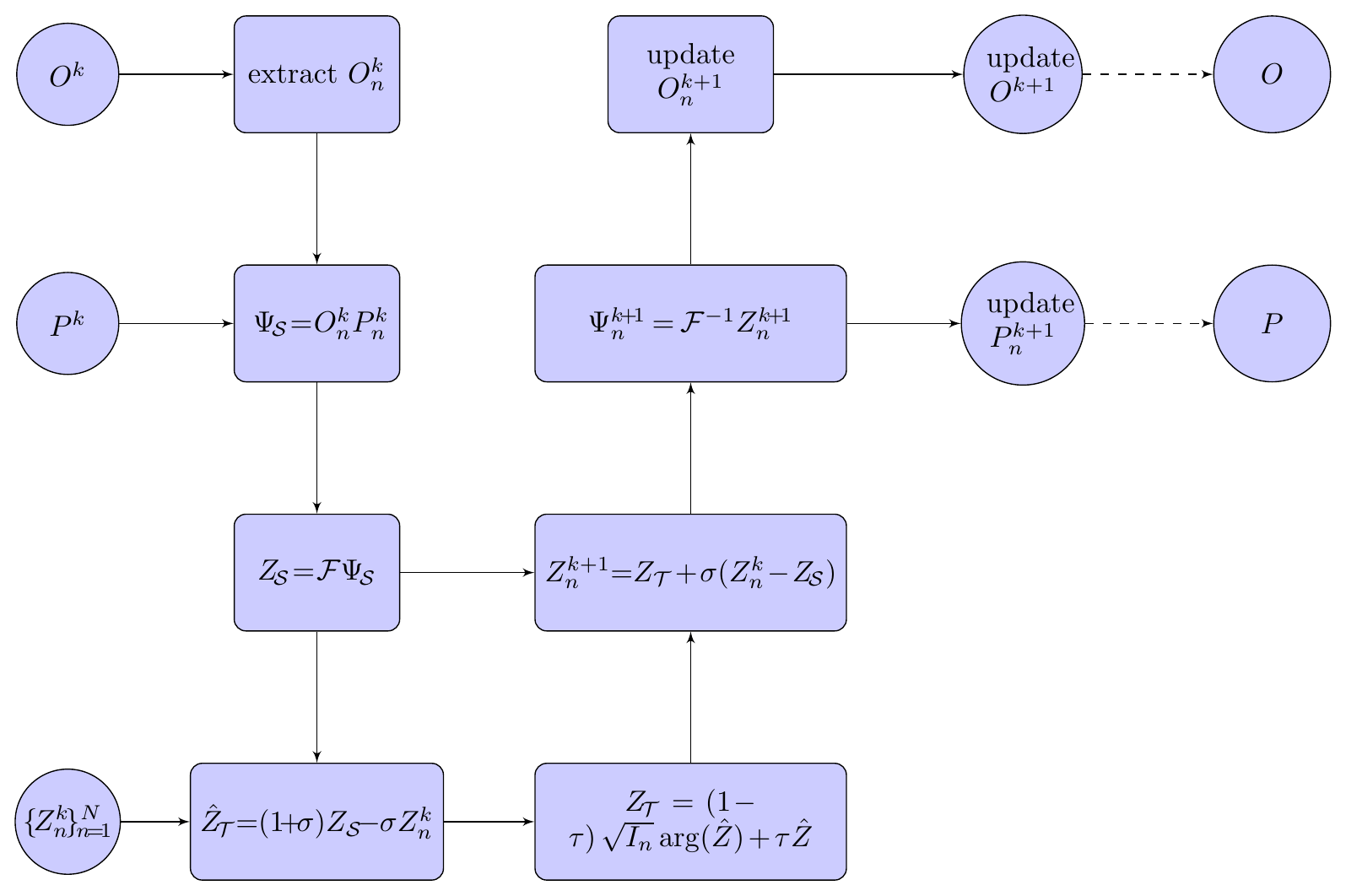}
    \caption{Flow chart of the sir-DR algorithm}
    \label{fig:flowchart}
\end{figure}

In combination of the semi-implicit algorithm and relaxed Douglas Rachford algorithm, we propose the sir-DR algorithm, shown in Fig. \ref{fig:flowchart}.

\begin{algorithm}
\caption{sir-DR algorithm}
\label{alg1}
\textbf{Input}: N measurements $\{I_n\}_{n=1}^N$,  number of iterations $K$, parameters $\sigma$, $\tau$, $\beta_O$, $\beta_P$\\
\textbf{Initialize}: $O^0$, $P^0$, $\{Z_n^0\}_{n=1}^N$.
\begin{algorithmic}
\FOR {$k = 1,\dots,K$}
\STATE randomly pick the $n^{th}$ diffraction pattern, extract $O_n^{k} = O^k|_{\Omega_n}$
\STATE update $\Psi_n^{k+1}$
\STATE \quad $\Psi_{\mathcal{S}} = O_n^k P^k$
\STATE \quad $Z_{\mathcal{S}} = \mathcal{F} \Psi_{\mathcal{S}} $
\STATE \quad $\hat{Z} = (1+\sigma) Z_{\mathcal{S}} - \sigma Z_n^k$
\STATE \quad $Z_{\mathcal{T}} = (1-\tau) \sqrt{I_n}  \arg{\hat{Z}} + \tau \hat{Z}$
\STATE \quad $Z_n^{k+1} = Z_{\mathcal{T}} + \sigma (Z_n^k- Z_{\mathcal{S}})$
\STATE \quad $\Psi_{n}^{k+1} = \mathcal{F}^{-1} Z_n^{k+1}$
\STATE update $O_n^{k+1}$, $P^{k+1}$
\STATE \quad $\displaystyle O_n^{k+1} = \frac{(1-\beta_O)\|P^k\|_{\max}^2 O_n^k + \beta_O \overline{P^k} \Psi_n^{k+1} }{(1-\beta_O)\|P^k\|_{\max}^2 + \beta_O |P^k|^2}$
\STATE \quad $\displaystyle P^{k+1} = P^k - \beta_P \frac{\overline{O_n^{k+1}} \big( P^k O_n^{k+1} - \Psi_n^{k+1} )}{\|O_n^{k+1}\|_{\max}^2} $
\STATE update $O^{k+1}$ 
\STATE \quad $O^{k+1}|_{\Omega_n} = O_n^{k+1}$
\ENDFOR
\end{algorithmic}
\textbf{Output}: $O^N$, $P^N$
\end{algorithm}

In this algorithm, we only apply the semi-implicit method on $O_n^k$ while $P^k$ can be integrated with the Forward Euler (gradient descent) method. $\tau\in[0,1]$ is chosen to be small. In most of our experiments, we select $\tau \approx 0.1$, while the choice of $\sigma$ depends on the specific problem. In many cases, $\sigma=1$ works very well (full reflection). But in some specific cases, large $\sigma$ might cause divergence or small recovered FOV. We decrease $\sigma$ in these cases, for example $\sigma=0.5$. We choose $\beta_O=0.9$ in most cases. $\beta_P$ is chosen to be large at the beginning ($\beta_P=1$) and decreases as a function of iteration. This adaptive step-size has been introduced as a strategy for noise-robust Fourier ptychography \cite{zuo2016}.

\section{Experimental results}
\subsection{Reconstruction from simulated data}
\begin{figure}
    \centering
    \begin{tikzpicture}
    \draw (0, 0) node[inner sep=0] {\includegraphics[height=4cm]{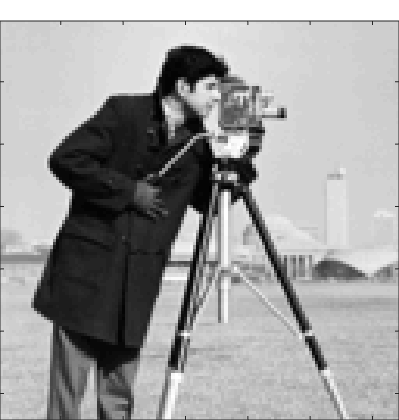}};
    \draw (-1.6, 1.4) node {\textcolor{black}{\textbf{\large a}}};
\end{tikzpicture}
    \begin{tikzpicture}
    \draw (0, 0) node[inner sep=0] {\includegraphics[height=4cm]{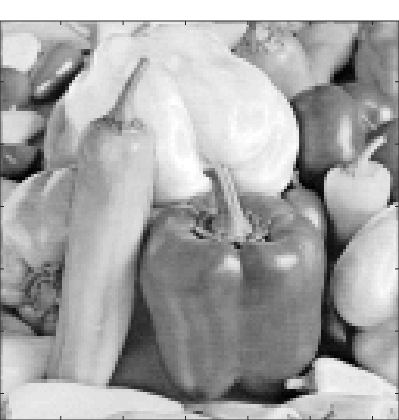}};
    \draw (-1.6, 1.4) node {\textcolor{black}{\textbf{\large b}}};
\end{tikzpicture}
    \caption{A simulated complex object with the amplitude being a camera man image (a) and the phase being a pepper image (b).}
    \label{fig:simudata}
\end{figure}

\begin{figure}
    \centering
    \begin{tikzpicture}
    \draw (0, 0) node[inner sep=0] {\includegraphics[height=4cm]{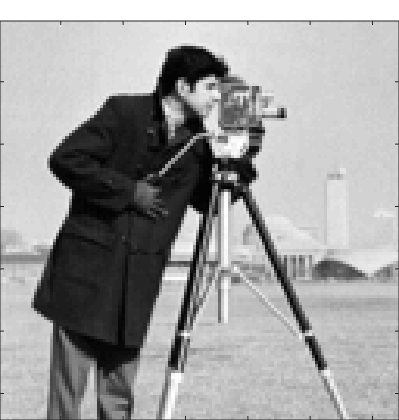}};
    \draw (-1.6, 1.5) node {\textcolor{black}{\textbf{\large a}}};
\end{tikzpicture}
    \begin{tikzpicture}
    \draw (0, 0) node[inner sep=0] {\includegraphics[height=4cm]{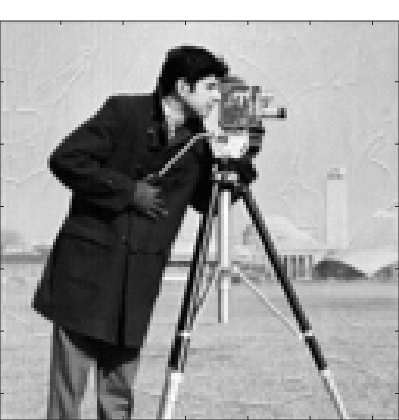}};
    \draw (-1.6, 1.5) node {\textcolor{black}{\textbf{\large b}}};
\end{tikzpicture}
    \begin{tikzpicture}
    \draw (0, 0) node[inner sep=0] {\includegraphics[height=4cm]{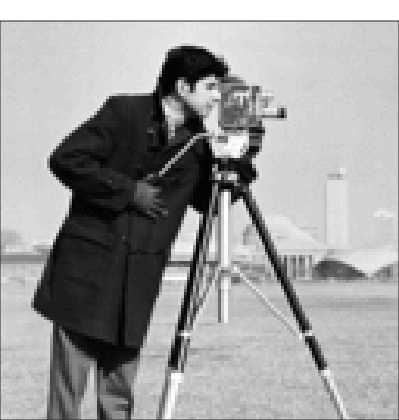}};
    \draw (-1.6, 1.5) node {\textcolor{black}{\textbf{\large c}}};
\end{tikzpicture} \\
    \hspace{0.5mm}
    \begin{tikzpicture}
    \draw (0, 0) node[inner sep=0] {\includegraphics[height=4cm]{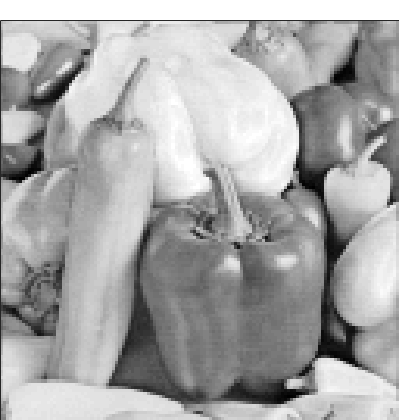}};
    \draw (-1.6, 1.5) node {\textcolor{black}{\textbf{\large d}}};
\end{tikzpicture}
    \begin{tikzpicture}
    \draw (0, 0) node[inner sep=0] {\includegraphics[height=4cm]{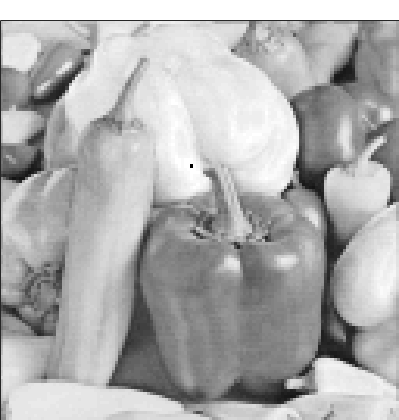}};
    \draw (-1.6, 1.5) node {\textcolor{black}{\textbf{\large e}}};
\end{tikzpicture}
    \begin{tikzpicture}
    \draw (0, 0) node[inner sep=0] {\includegraphics[height=4cm]{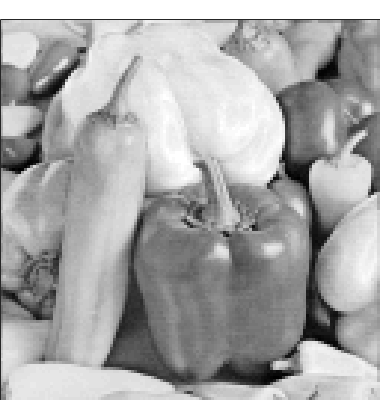}};
    \draw (-1.6, 1.5) node {\textcolor{black}{\textbf{\large f}}};
\end{tikzpicture}
    
    \caption{The reconstructions of ePIE, rPIE and sir-DR of a complex object consisting of $128\times128$ pixels, a scan step size of 35 pixels and $4\times4$ diffraction patterns. Poisson noise was added to the diffraction patterns with $R_{noise} = 3.73\%$. Top row (a-c) shows the amplitude and bottom row (d-f) shows the phase of ePIE, rPIE and sir-DR reconstructions, respectively.}
    \label{fig:camera}
\end{figure}
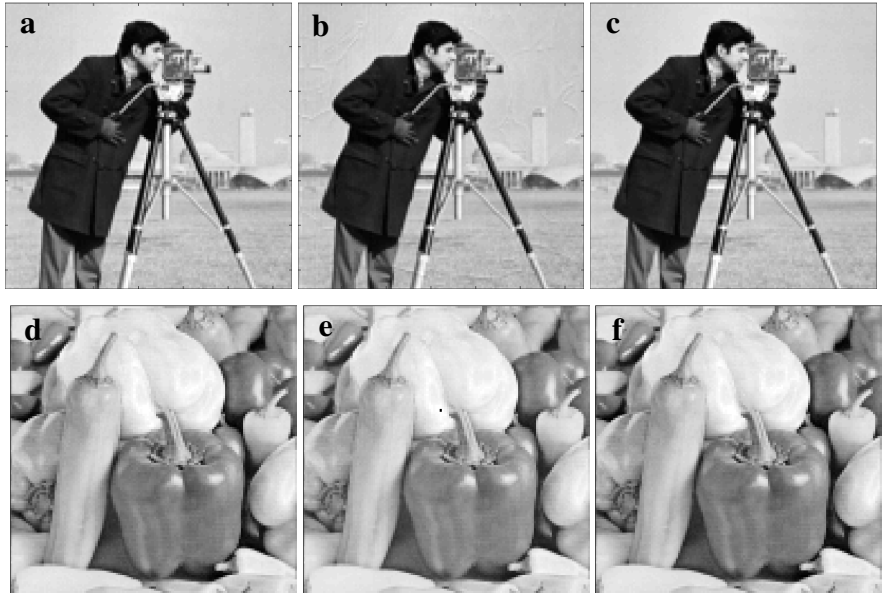

To examine the sir-DR algorithm, we simulate a complex object of $128 \times 128$ pixels with a cameraman and a pepper images representing the amplitude and the phase, respectively (Fig. \ref{fig:simudata}).The circular aperture is chosen as probe with a radius of 50 pixels. We raster scan the aperture over the object with a step size of 35 pixels, resulting in 4x4 scan positions. The overlap is therefore 56.4\%, the approximate lower limit for ePIE to work. Poisson noise is added to the diffraction patterns with a  flux of $10^8$ photons per scan position. We use $R_{noise}$ to quantify the relative error with respect to the noise-free diffraction patterns
\begin{align}
    R_{noise} = \frac{1}{N} \sum_{n=1}^N \frac{ \| |\mathcal{F} ( P^0 O^0_n ) | - \sqrt{I_n}  \|_{1,1} }{ \| \sqrt{I_n} \|_{1,1}}.
\end{align}
where $P^0$ and $O^0$ is the noise-free model and the $L_{1,1}$ matrix norm represents the sum of all elements in absolute value of the matrix. The above flux results in  $R_{noise} = 3.73\%$. Fig. \ref{fig:camera} shows that three algorithms (ePIE, rPIE and sir-DR) all successfully reconstruct the object in the case where the overlap between adjacent positions is high and the noise level is low.

As a baseline comparison, Fig. \ref{fig:camera} shows that all three algorithms correctly reconstruct the object in the ideal case when the overlap between adjacent positions is high and the noise level is low.

\begin{figure}
    \centering
    \begin{tikzpicture}
    \draw (0, 0) node[inner sep=0] {\includegraphics[height=4cm]{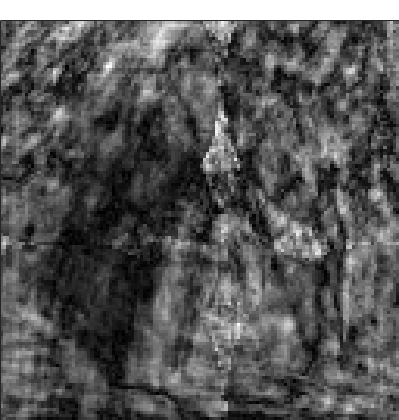}};
    \draw (-1.6, 1.5) node {\textcolor{white}{\textbf{\large a}}};
    \end{tikzpicture}
    \begin{tikzpicture}
    \draw (0, 0) node[inner sep=0] {\includegraphics[height=4cm]{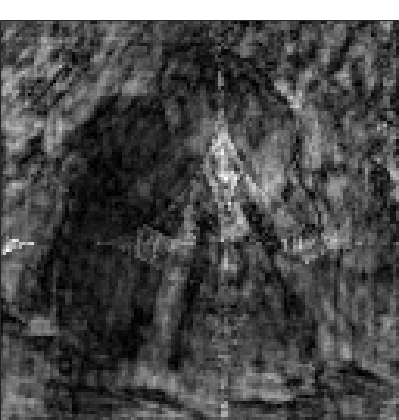}};
    \draw (-1.6, 1.5) node {\textcolor{white}{\textbf{\large b}}};
    \end{tikzpicture}
    \begin{tikzpicture}
    \draw (0, 0) node[inner sep=0] {\includegraphics[height=4cm]{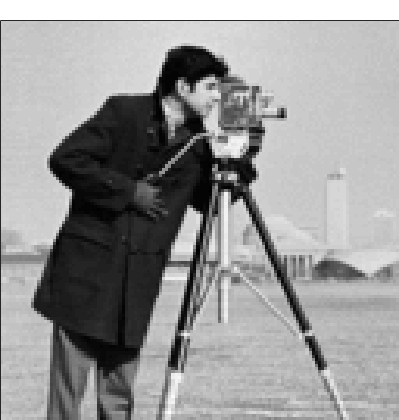}};
    \draw (-1.6, 1.5) node {\textcolor{black}{\textbf{\large c}}};
    \end{tikzpicture} 
     
    \begin{tikzpicture}
    \draw (0, 0) node[inner sep=0] {\includegraphics[height=4cm]{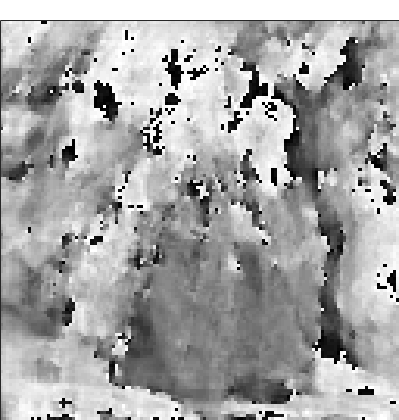}};
    \draw (-1.6, 1.5) node {\textcolor{black}{\textbf{\large d}}};
    \end{tikzpicture}
    \begin{tikzpicture}
    \draw (0, 0) node[inner sep=0] {\includegraphics[height=4cm]{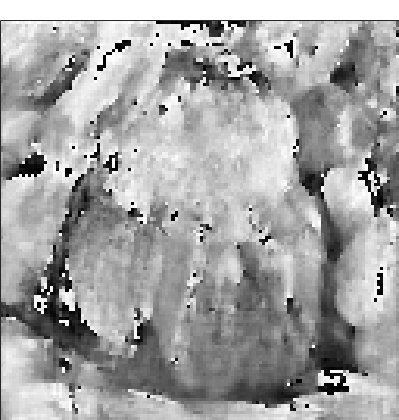}};
    \draw (-1.6, 1.5) node {\textcolor{black}{\textbf{\large e}}};
    \end{tikzpicture}
    \begin{tikzpicture}
    \draw (0, 0) node[inner sep=0] {\includegraphics[height=4cm]{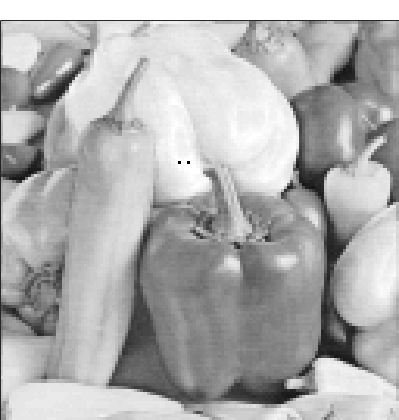}};
    \draw (-1.6, 1.5) node {\textcolor{black}{\textbf{\large f}}};
    \end{tikzpicture}

    \caption{Ptychographic reconstructions of sparse data by ePIE, rPIE and sir-DR. The data consist of $3\times3$ diffraction patterns with a scan step of 50 pixels.  Poisson noise was added to the diffraction patterns with $R_{noise} = 3.73\%$. Top row (a-c) shows the amplitude and bottom row (d-f) shows the phase of ePIE, rPIE and sir-DR reconstructions, respectively. For this sparse data, ePIE and rPIE fail to converge no matter how many iterations are used, but sir-DR converges to an image of good quality.}
    \label{fig:camera_sparse}
\end{figure}

Next, we apply the three algorithms to the reconstruction of sparse data, which is centrally important to reducing computation time, data storage requirements and incident dose to the sample. We increase the scan step size to 50 pixels while keeping the same field of view, which reduces the number of diffraction patterns to $3\times3$. Consequently, the overlap is reduced to 39.1\%. Not only is the overlap between adjacent positions low, but the total number of measurements is also small, creating a challenging data set for conventional ptychographic algorithms. Fig. \ref{fig:camera_sparse} show that sir-DR can work well with sparse data, while ePIE and rPIE fail to reconstruct the object faithfully.

\subsection{Reconstruction from experimental data}
\subsubsection{Optical laser data}
As an initial test of sir-DR with experimental data, we collect diffraction patterns from an USAF resolution pattern using a green laser with a wavelength of 543 nm. The incident illumination is created by a $15\, \mu m$ diameter pinhole. The pinhole is placed approximately 6 mm in front of the sample, creating a illumination wavefront on the sample plane that can be approximated by Fresnel propagation. The detector is positioned 26 cm downstream of the sample to collect far-field diffraction patterns. We raster scan across the sample with a step size of $50\, \mu m$ and acquire 169 diffraction patterns. We perform a sparsity test by randomly choosing 85 diffraction patterns (50\% density) and run ePIE, rPIE, and sir-DR on this subset with 300 iterations. If we assume the probe diameter is to where the intensity falls to 10\% of the maximum, then the the overlaps are 73\% and  46.4\% for the full and sparsity sets respectively. Fig. 5 shows that rPIE and sir-DR obtain a larger FOV than ePIE as both use regularization. Furthermore, sir-DR removes noise more effectively and obtains a flatter background than ePIE and rPIE. We monitor the R-factor (relative error) to quantify the reconstruction, defined as
\begin{align}
    R_F = \frac{1}{N} \sum_{n=1}^N \frac{ \| |\mathcal{F} ( P O_n ) | - \sqrt{I_n} \|_{1,1}}{ \| \sqrt{I_n} \|_{1,1}}
\end{align}
$R_F$  is 16.94\%, 13.95\% and 13.28\% for the ePIE, rPIE and sir-DR reconstructions, respectively. 
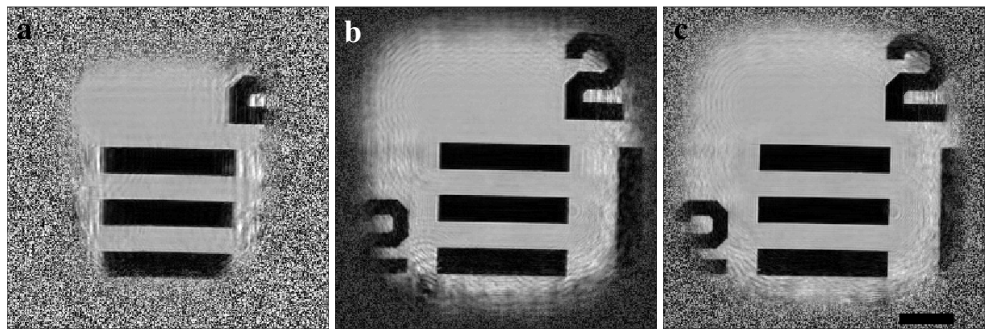
\begin{figure}
    \centering
    \begin{tikzpicture}
    \draw (0, 0) node[inner sep=0] {\includegraphics[height=4.5cm]{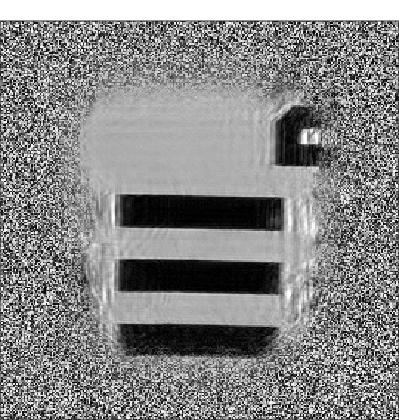}};
    \draw (-1.9, 1.7) node {\textcolor{black}{\textbf{\large a}}};
    \end{tikzpicture}
    \begin{tikzpicture}
    \draw (0, 0) node[inner sep=0] {\includegraphics[height=4.5cm]{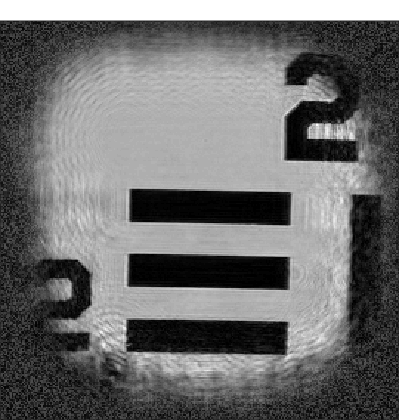}};
    \draw (-1.9, 1.7) node {\textcolor{white}{\textbf{\large b}}};
    \end{tikzpicture}
    \begin{tikzpicture}
    \draw (0, 0) node[inner sep=0] {\includegraphics[height=4.5cm]{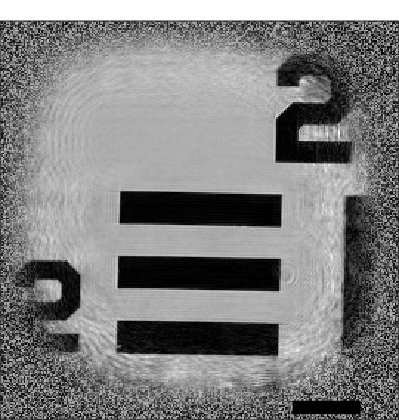}};
    \draw (-1.9, 1.7) node {\textcolor{black}{\textbf{\large c}}};
    \end{tikzpicture}
    \caption{The ePIE (a), rPIE (b) and sir-DR (c) reconstructions of a sparse data with 300 iterations, where sir-DR obtains a better quality reconstruction than ePIE and rPIE. Both sir-DR and rPIE produce a larger FOV than ePIE. Scale bar $200 \mu m$}
    \label{fig:USAF}
\end{figure}

\subsubsection{Synchrotron radiation data}
\begin{figure}[http]
    \centering
    \hspace{-2cm}
    \begin{subfigure}[b1]{0.2\textwidth}
    \begin{tikzpicture}
    \draw (0, 0) node[inner sep=0] {\includegraphics[width=4cm]{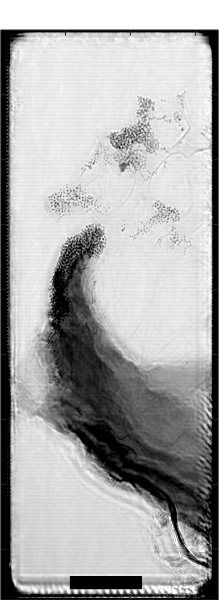}};
    \draw (-1.6, 4.6) node {\textcolor{black}{\textbf{\large a}}};
    \end{tikzpicture}
    \end{subfigure}\hspace{1.5cm}
    \begin{subfigure}[b2]{0.2\textwidth}
    \begin{tikzpicture}
    \draw (0, 0) node[inner sep=0] {\includegraphics[width=4.9cm]{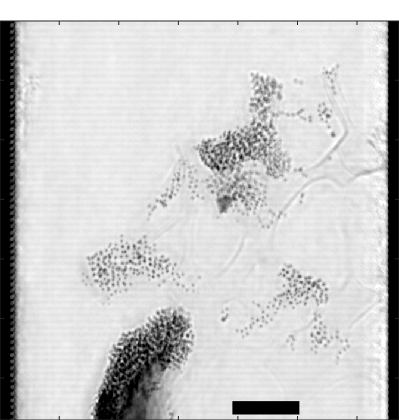}};
    \draw (-2.1, 2.0) node {\textcolor{black}{\textbf{\large b}}};
    \end{tikzpicture}
    \begin{tikzpicture}
    \draw (0, 0) node[inner sep=0] {\includegraphics[width=4.9cm]{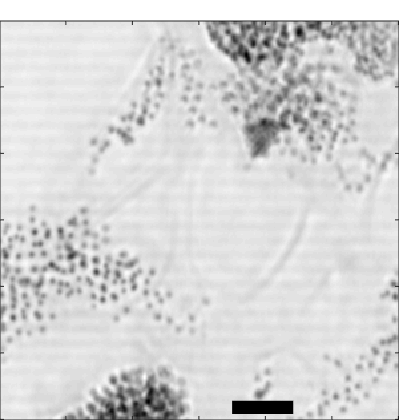}};
    \draw (-2.1, 2.0) node {\textcolor{black}{\textbf{\large c}}};
    \end{tikzpicture}
    \end{subfigure}
    \caption{The ePIE reconstruction of a portion of a HeLa cell labeled with nanoparticles after 300 iterations. The data consist of 7,500 diffraction patterns and covers a $9.71\times 3.70\, \mu m$ region. (a) The full FOV of the sample. (b) Magnified view of a region ($3.70\times 3.70 \, \mu m$) in (a). (c) Magnified view of a region ($1.66 \times 1.66 \mu m$) in (b). The scale bars are $1000nm, \,  500 nm$ and $200nm$ respectively.}
    \label{fig:ALS_0}
\end{figure}
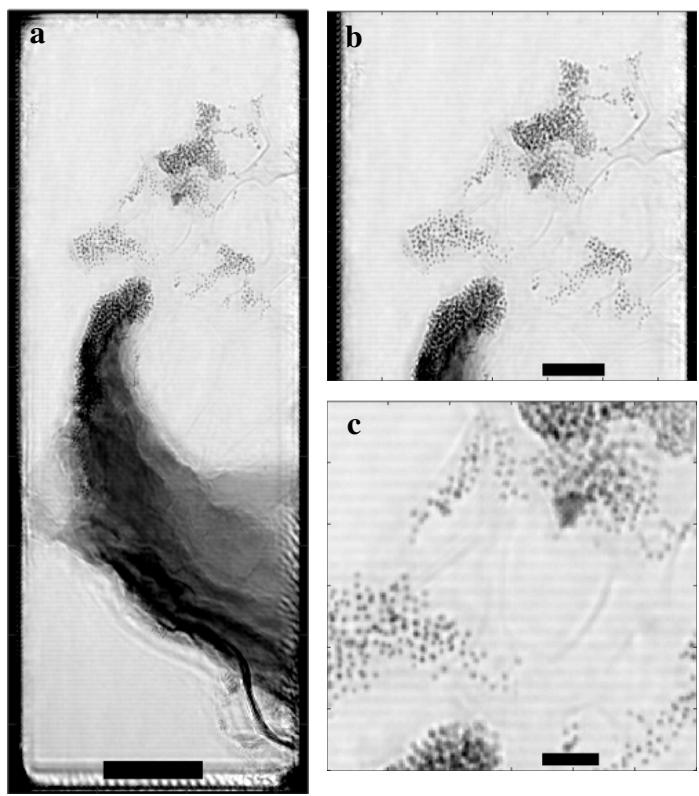

To demonstrate the applicability of sir-DR to synchrotron radiation data, we reconstruct a ptychographic data set collected from the Advanced Light Sources  \cite{Gallagher2017}. In this experiment, 710 eV soft x-rays are focused onto a sample using a zone plate and the far-field diffraction patterns are collected by a detector. A 2D scan consists of 7,500 positions, which span approximately $10\times4\, \mu m $. The sample is a portion of a HeLa cell labeled with nanoparticles, which is supported on a graphene-oxide layer. Fig. \ref{fig:ALS_0} shows the ePIE reconstruction of the whole FOV of the sample. To compare the three algorithms, we choose a subdomain of a $4\times4\, \mu m$ region, consisting of 2,450 diffraction patterns. With the same assumption, the overlap is computed to be 79.5\%. 

\begin{figure}
    \centering
    \begin{tikzpicture}
    \draw (0, 0) node[inner sep=0] {\includegraphics[height=4.4cm]{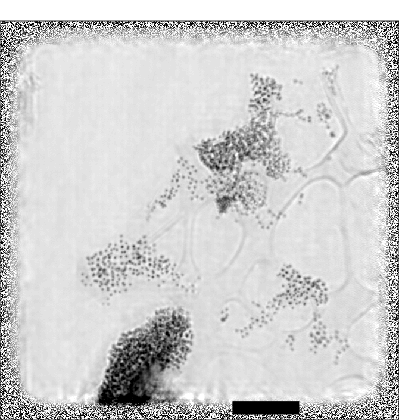}};
    \draw (-1.8, 1.6) node {\textcolor{black}{\textbf{\large a}}};
    \end{tikzpicture}
    \begin{tikzpicture}
    \draw (0, 0) node[inner sep=0] {\includegraphics[height=4.4cm]{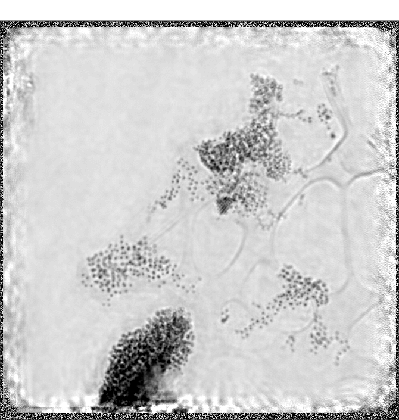}};
    \draw (-1.8, 1.6) node {\textcolor{black}{\textbf{\large b}}};
    \end{tikzpicture}
    \begin{tikzpicture}
    \draw (0, 0) node[inner sep=0] {\includegraphics[height=4.4cm]{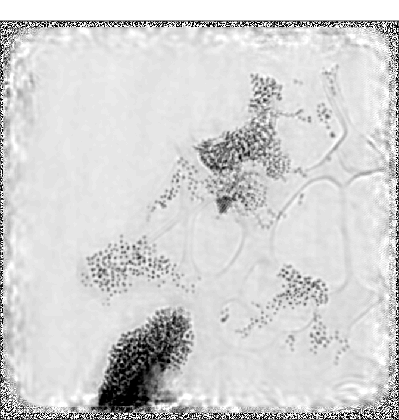}};
    \draw (-1.8, 1.6) node {\textcolor{black}{\textbf{\large c}}};
    \end{tikzpicture}    

    \begin{tikzpicture}
    \draw (0, 0) node[inner sep=0] {\includegraphics[height=4.4cm]{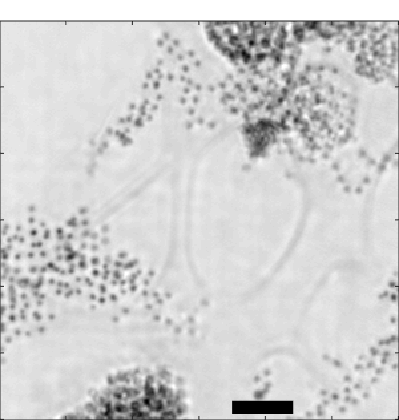}};
    \draw (-1.8, 1.6) node {\textcolor{black}{\textbf{\large d}}};
    \end{tikzpicture}
    \begin{tikzpicture}
    \draw (0, 0) node[inner sep=0] {\includegraphics[height=4.4cm]{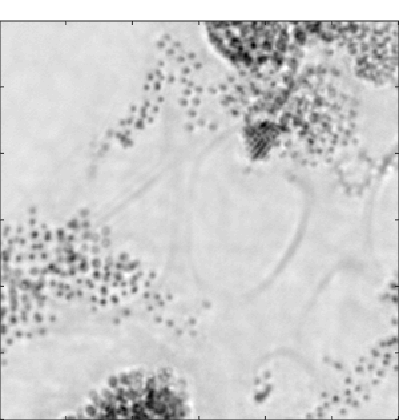}};
    \draw (-1.8, 1.6) node {\textcolor{black}{\textbf{\large e}}};
    \end{tikzpicture}
    \begin{tikzpicture}
    \draw (0, 0) node[inner sep=0] {\includegraphics[height=4.4cm]{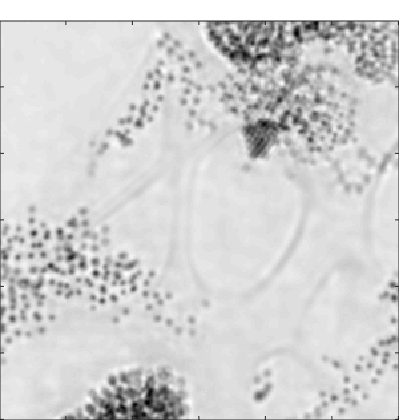}};
    \draw (-1.8, 1.6) node {\textcolor{black}{\textbf{\large f}}};
    \end{tikzpicture} 
    
    \begin{tikzpicture}
    \draw (0, 0) node[inner sep=0] {\includegraphics[height=4.4cm]{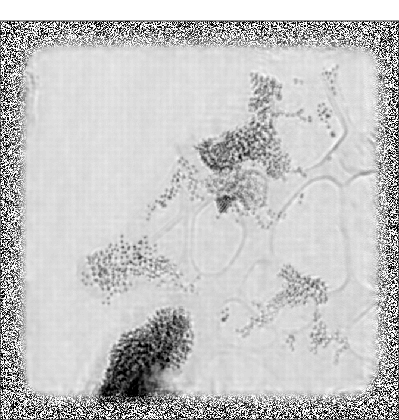}};
    \draw (-1.8, 1.6) node {\textcolor{black}{\textbf{\large g}}};
    \end{tikzpicture}
    \begin{tikzpicture}
    \draw (0, 0) node[inner sep=0] {\includegraphics[height=4.4cm]{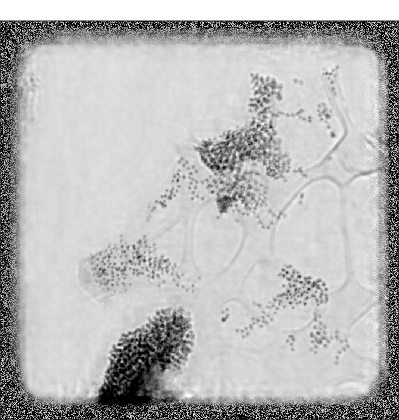}};
    \draw (-1.8, 1.6) node {\textcolor{black}{\textbf{\large h}}};
    \end{tikzpicture}
    \begin{tikzpicture}
    \draw (0, 0) node[inner sep=0] {\includegraphics[height=4.4cm]{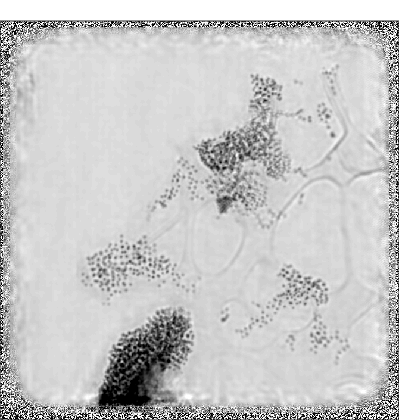}};
    \draw (-1.8, 1.6) node {\textcolor{black}{\textbf{\large i}}};
    \end{tikzpicture}     
    
    \begin{tikzpicture}
    \draw (0, 0) node[inner sep=0] {\includegraphics[height=4.4cm]{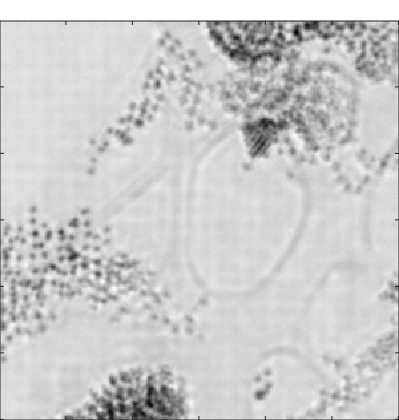}};
    \draw (-1.8, 1.6) node {\textcolor{black}{\textbf{\large j}}};
    \end{tikzpicture}
    \begin{tikzpicture}
    \draw (0, 0) node[inner sep=0] {\includegraphics[height=4.4cm]{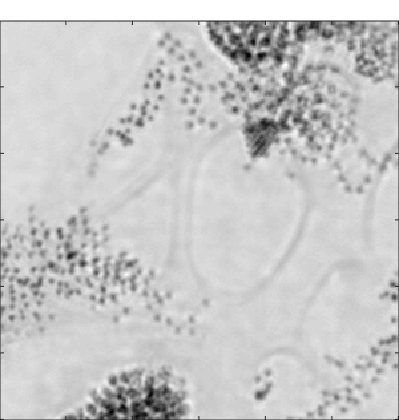}};
    \draw (-1.8, 1.6) node {\textcolor{black}{\textbf{\large k}}};
    \end{tikzpicture}
    \begin{tikzpicture}
    \draw (0, 0) node[inner sep=0] {\includegraphics[height=4.4cm]{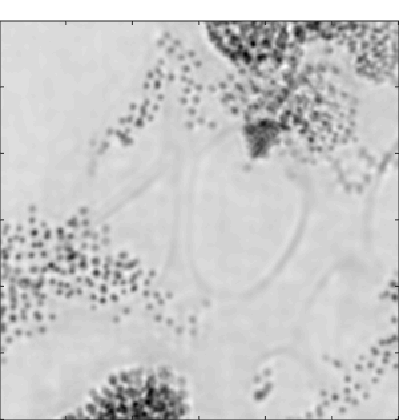}};
    \draw (-1.8, 1.6) node {\textcolor{black}{\textbf{\large l}}};
    \end{tikzpicture}   
    \caption{The ePIE (a), rPIE (b) and sir-DR (c) reconstructions of a $3.70 \times 3.70 \mu m$ region of the HeLa cell after 300 iterations with $R_F = 16.48\%, \, 16.50\%$ and $14.40\%$, respectively. (d-f) Magnified regions ($1.66 \times 1.66 \mu m$) in (a-c), respectively. (g-i) ePIE, rPIE and sir-DR reconstructions after 100 with $R_F = 18.69\%, \, 16.60\%$ and $14.92\%$, respectively. (j-l) Magnified regions in (g-i), respectively. Among the three algorithms, sir-DR not only converges the fastest, but also produces the best reconstruction. Scale bar $500 nm$ and $200nm$ respectively.}
    \label{fig:ALS}
\end{figure}
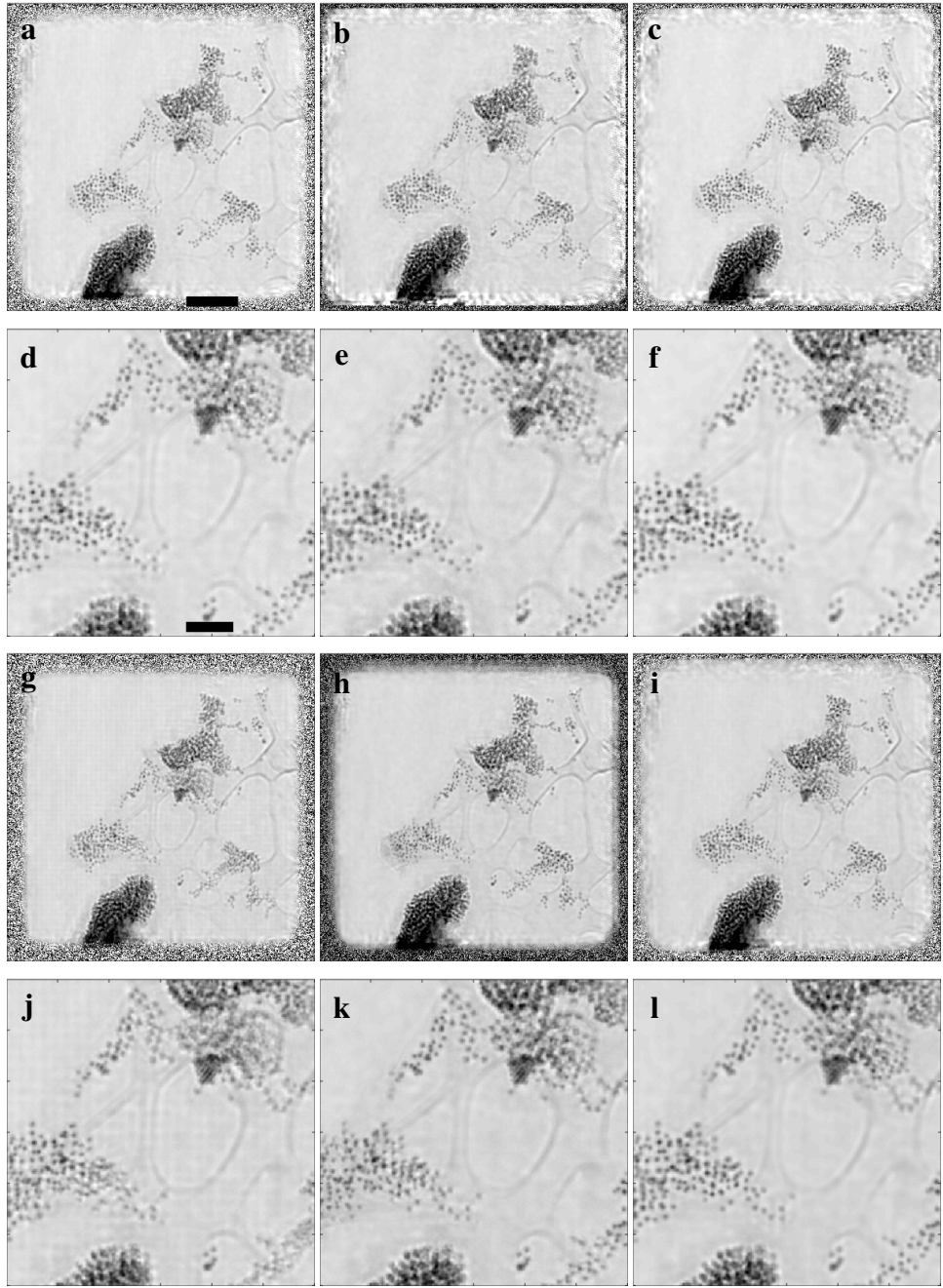

Fig. \ref{fig:ALS} show the ePIE, rPIE, and sir-DR reconstructions, respectively. With 300 iterations, all three algorithms converge to images with good quality. When reducing the number of iterations to 100, we observe that sir-DR converges faster and reconstruct a larger FOV than ePIE. The individual nanoparticles, which serve as a resolution benchmark, are better resolved in the sir-DR reconstruction than the ePIE and rPIE ones. Furthermore, the reconstruction by ePIE contains artifacts as a faint square grid, which is removed by rPIE and sir-DR.

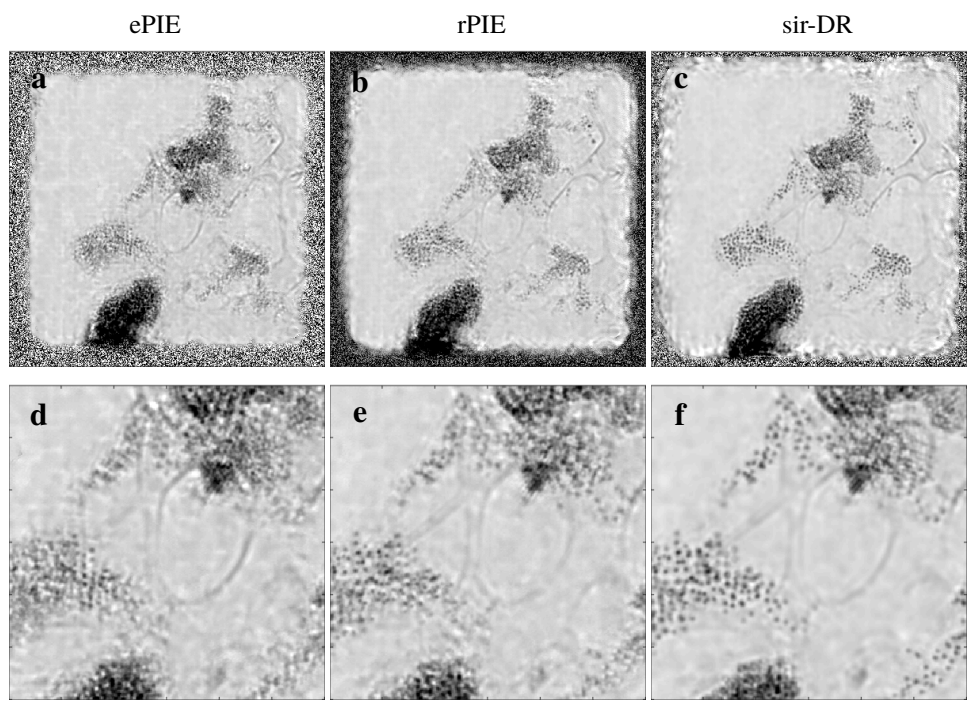
\begin{figure}
    \centering
    \hspace{0cm} ePIE \hspace{3.5cm} rPIE \hspace{3.5cm} sir-DR \\
    \begin{tikzpicture}
    \draw (0, 0) node[inner sep=0] {\includegraphics[height=4.4cm]{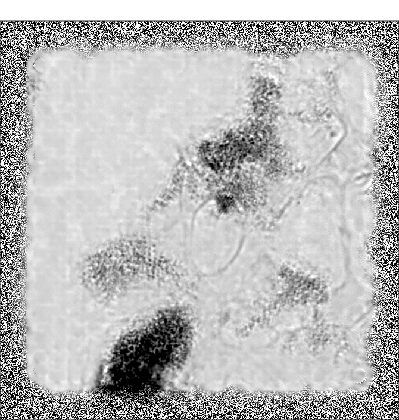}};
    \draw (-1.7, 1.6) node {\textcolor{black}{\textbf{\large a}}};
    \end{tikzpicture}
    \begin{tikzpicture}
    \draw (0, 0) node[inner sep=0] {\includegraphics[height=4.4cm]{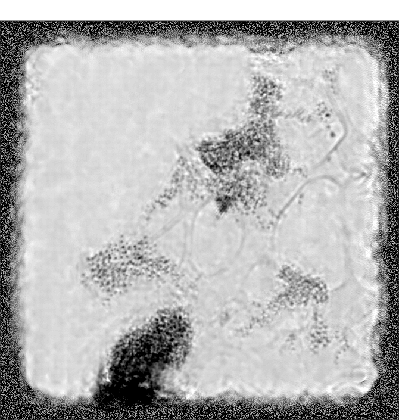}};
    \draw (-1.7, 1.6) node {\textcolor{black}{\textbf{\large b}}};
    \end{tikzpicture}
    \begin{tikzpicture}
    \draw (0, 0) node[inner sep=0] {\includegraphics[height=4.4cm]{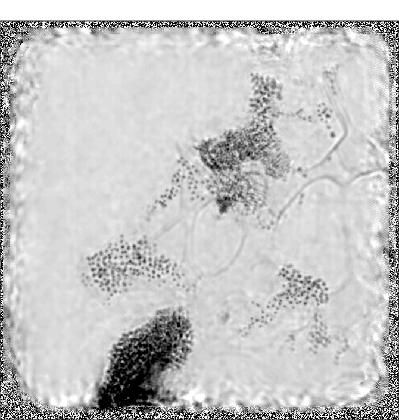}};
    \draw (-1.7, 1.6) node {\textcolor{black}{\textbf{\large c}}};
    \end{tikzpicture} 
    
    \begin{tikzpicture}
    \draw (0, 0) node[inner sep=0] {\includegraphics[height=4.4cm]{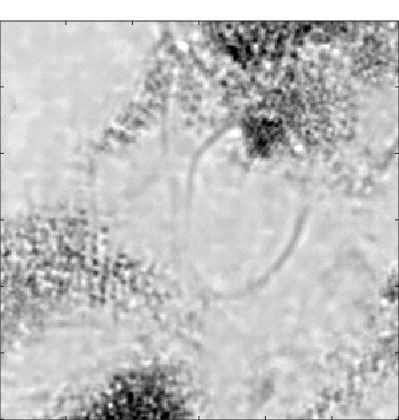}};
    \draw (-1.7, 1.6) node {\textcolor{black}{\textbf{\large d}}};
    \end{tikzpicture}
    \begin{tikzpicture}
    \draw (0, 0) node[inner sep=0] {\includegraphics[height=4.4cm]{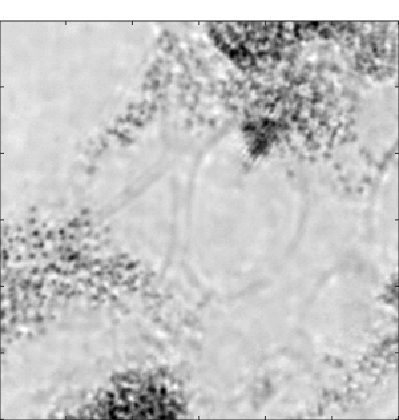}};
    \draw (-1.7, 1.6) node {\textcolor{black}{\textbf{\large e}}};
    \end{tikzpicture}
    \begin{tikzpicture}
    \draw (0, 0) node[inner sep=0] {\includegraphics[height=4.4cm]{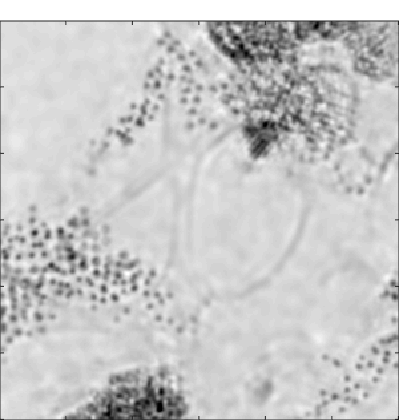}};
    \draw (-1.7, 1.6) node {\textcolor{black}{\textbf{\large f}}};
    \end{tikzpicture}    
    
    \caption{(a-c) The ePIE, rPIE, and sir-DR reconstructions of a sparse data set after 300 iterations with $R_F = 19.52\%, \,16.58\%$ and $14.26\%$, respectively. To create the sparse data, we randomly pick 980 out of 2,450 diffraction patterns from the HeLa sample. (d-f) Magnified regions in (a-c), respectively. While the quality of the ePIE and rPIE reconstruction is degraded, sir-DR reproduces a good quality image with more distinguishable features.}
    \label{fig:ALS_sparse}
\end{figure}

We next perform a sparsity test by randomly picking 980 out of 2,450 diffraction patterns, i.e. a reduction of data by 60\%. The corresponding overlap of the sparsity set is 50.8\%. Fig. \ref{fig:ALS_sparse} shows the reconstructions by ePIE, rPIE, and sir-DR with 300 iterations. Both the ePIE and rPIE reconstructions exhibit noticeable degradation. In particular, the nanoparticles are not well resolved. But the sir-DR reconstruction has no noticeable artifacts noise and the individual nanoparticles are clearly visible. The quality of sir-DR reconstruction with 60\% data reduction is still comparable to that of ePIE using all the diffraction patterns

\section{Conclusion}
In this work, we have developed a fast and robust ptychographic algorithm, termed sir-DR. The algorithm relaxes Douglas-Rachford to improve robustness and applies a semi-implicit scheme (semi-Backward Euler) to solve for the object and to expand the reconstructed FOV. Using both simulated and experimental data, we have demonstrated that sir-DR outperforms ePIE and rPIE with sparse data. Being able to obtain good ptychographic reconstructions from sparse measurements, sir-DR can reduce the computation time, data storage requirement and radiation dose to the sample.

\section*{Acknowledgments}
This work was supported by STROBE: A National Science Foundation Science \& Technology Center (DMR-1548924). J.M. also acknowledges the support by the Office of Basic Energy Sciences of the US DOE (DE-SC0010378). 



\bibliography{sample}






\end{document}